\documentclass[graybox]{svmult}

\usepackage{subfigure}
\usepackage{float}
\usepackage{mathptmx}       
\usepackage{helvet}         
\usepackage{courier}        
\usepackage{type1cm}        
\usepackage{makeidx}         
\usepackage{graphicx}        
\usepackage{multicol}        
\usepackage[bottom]{footmisc}

\makeindex             

\begin{document}

\title*{Identifying Gene-Environment Interactions with A Least Relative Error Approach}
\author{Yangguang Zang, Yinjun Zhao, Qingzhao Zhang, Hao Chai, Sanguo Zhang and Shuangge Ma$^{*}$}
\titlerunning{Identifying G$\times$E Interactions with A Least Relative Error Approach}
\institute{Yangguang Zang, Sanguo Zhang
 \at School of Mathematical Sciences, University of Chinese Academy of Sciences\\Key Laboratory of Big Data Mining and Knowledge Management, Chinese Academy of Sciences, E-mail:zangyangguang@mails.ucas.ac.cn, sgzhang@ucas.ac.cn
\and Yinjun Zhao, Qingzhao Zhang, Hao Chai, Shuangge Ma (*Corresponding author) \at Department of Biostatistics, Yale University, E-mail:yinjun.zhao@yale.edu, qzzhang@ucas.ac.cn, hao.chai@yale.edu, shuangge.ma@yale.edu}
%
%
\maketitle

\abstract{For complex diseases, the interactions between genetic and environmental risk factors can have important implications beyond the main effects. Many of the existing interaction analyses conduct marginal analysis and cannot accommodate the joint effects of multiple main effects and interactions. In this study, we conduct joint analysis which can simultaneously accommodate a large number of effects. Significantly different from the existing studies, we adopt loss functions based on relative errors, which offer a useful alternative to the ``classic" methods such as the least squares and least absolute deviation. Further to accommodate censoring in the response variable, we adopt a weighted approach. Penalization is used for identification and regularized estimation. Computationally, we develop an effective algorithm which combines the majorize-minimization and coordinate descent. Simulation shows that the proposed approach has satisfactory performance. We also analyze lung cancer prognosis data with gene expression measurements.}

\section{Introduction}
For complex diseases, it is of significant interest to identify genetic risk factors. For etiology, biomarkers, and prognosis, the interactions between genetic and environmental risk factors, also referred to as G $\times$ E interactions, have important implications beyond the main effects. Extensive studies have been conducted to search for important G $\times$ E interactions \cite{cordell2009detecting,hunter2005gene,north2008importance,wu2014selective}. In this article, we focus on analyzing prognosis, which has an essential role in biomedical studies. It is conjectured that the proposed approach can be extended to some types of disease outcomes.

Denote $T$ as the survival time, $Z =(Z_{1},\cdots,Z_{p})^\top\in R^{p\times 1}$ as the $p$ genetic factors (G), and $X = (X_{1},\cdots,X_{q})^\top \in R^{q\times1}$ as the $q$ clinical/environmental risk factors (E). There are two families of methods for detecting the main G and E effects and G$\times$E interactions. The first conducts marginal analysis and analyzes one or a small number of G factors at a time \cite{hunter2005gene,shi2014penalized,thomas2010methods}. With a slight abuse of terminology, we use the generic phrase ``gene" for the G factor. In marginal analysis, for gene $k$, consider the model $T\sim \phi(\sum_{j=1}^qX_j\alpha_j+Z_k\beta_k+\sum_{j=1}^q X_jZ_k\xi_{jk})$, where model $\phi$ is known up to the regression coefficients, and $\alpha_j,\beta_k,\xi_{jk}$ are the unknown regression coefficients. Marginal analysis is easy to implement however cannot accommodate the joint effects of multiple main effects and interactions. The second family conducts joint analysis and describes the joint effects of all factors in a single model \cite{liu2013identification, wu2014integrative, zhu2014identifying}. More specifically, consider  $T\sim \phi(\sum_{j=1}^qX_j\alpha_j+\sum_{k=1}^pZ_k\beta_k+\sum_{j=1}^q\sum_{k=1}^p X_jZ_k\xi_{jk})$.

For the simplicity of description, consider the simple linear regression setting. In the literature, most of the existing methods adopt loss functions built on {\it absolute error criteria}, with the most popular including the least squares (LS) and least absolute deviation (LAD). Under certain settings, it has been found that the {\it relative errors} are more sensible \cite{khoshgoftaar1992predicting, park1998relative, van2015relative}. The most distinguishable feature of the relative error-based approaches is that they are scale-free, which, as discussed in the published studies \cite{chen2010least}, can be advantageous in survival and other analysis.
There are at least two ways of defining relative error-based criteria. The first is defined based on the ratio of the error with respect to the target. The second is defined on the ratio of the error with respect to the predictor \cite{chen2010least}. Based on the two types of relative errors, researchers have proposed the least absolute relative errors (LARE) criterion and the least product relative errors (LPRE) criterion for linear multiplicative models. The LARE criterion is convex but not smooth. For its extensions and applications, we refer to \cite{li2014empirical, tsionas2014bayesian, zhang2013local} and followup studies. In comparison, the LPRE criterion is smooth and convex \cite{chen2013least}. Under low-dimensional settings, asymptotical properties of the LARE and LPRE estimates for linear multiplicative models have been established \cite{chen2010least, chen2013least}.

Different from the existing ones (which focus on the main effects), this study adopts the relative error-based criteria for analyzing interactions. Such new criteria may provide a useful alternative to the commonly-adopted absolute error-based criteria. In genetic data analysis, it is critical to identify the important main effects and interactions, which poses a variable (model) selection problem. In two recent studies \cite{xia2015regularized, zhang2013local}, variable selection based on the LARE has been studied. However, the existing studies are limited to the situation where the dimension of model is smaller than the sample size. To the best of our knowledge, there is a lack of study examining the relative error-based criteria under high-dimensional settings. Also different from the existing studies, we analyze prognosis data under right censoring, which introduces additional complexity. 

\section{Methods}

\subsection{Model and relative error-based criteria}

For modeling a prognosis response, we consider the following linear multiplicative (accelerated failure time - AFT) model,
\begin{equation}\label{AFT}
T=\exp\Big(\sum_{j=1}^qX_j\alpha_j+\sum_{k=1}^pZ_k\beta_k+\sum_{j=1}^q\sum_{k=1}^p X_jZ_k\xi_{jk}\Big)\varepsilon,
\end{equation}
where $\varepsilon$ is the random error independent of $X$ and $Z$. This model provides a useful alternative to the Cox and other models. It can be especially preferred under high-dimensional settings. Let $U=(X^\top, Z^\top, (X\otimes Z)^\top)^\top$ and $\theta=(\alpha^\top, \beta^\top,\xi^\top)^\top$, then we can write model (\ref{AFT}) as
\begin{equation}\label{AFT2}
T=\exp(U^\top\theta)\varepsilon.
\end{equation}

First consider the case without censoring. Suppose that we have $n$ iid observations $\{t_{i},\textbf{x}_{i},\textbf{z}_{i}\}_{i=1}^n$, where $\textbf{x}_{i} = (x_{i1}, \cdots, x_{iq})^\top$ and $\textbf{z}_{i} = (z_{i1}, \cdots, z_{ip})^\top$. Denote $\textbf{u}_{i} = (\textbf{x}_{i}^\top, \textbf{z}_{i}^\top, (\textbf{x}_{i}\otimes \textbf{z}_{i})^\top)^\top$.  With the logarithm transformation, model (\ref{AFT2}) can be rewritten as $\log(T)=U^\top\theta+\log(\varepsilon)$. The LS and LAD methods can be applied, which, respectively, minimize the objective functions $\sum_{i=1}^n (\log(t_i)-\textbf{u}_i^\top\theta)^2$ and $\sum_{i=1}^n |\log(t_i)-\textbf{u}_i^\top\theta|$. Both methods are built on the absolute errors.

As discussed in the literature, under certain scenarios, the relative error-based criteria can be more sensible. In this article, we consider the least absolute relative errors (LARE) \cite{chen2010least} and least product relative errors (LPRE) \cite{chen2013least} criteria. They have been relatively more popular in the relative error literature and deserve a higher priority.  The LARE objective function is defined as
\begin{equation}\label{LARE}
LARE_n(\theta)=\sum_{n=1}^n \left\{\left|\frac{t_i-\exp(\textbf{u}_i^\top \theta)}{t_i}\right| + \left|\frac{t_i-\exp(\textbf{u}_{i}^\top \theta)}{\exp(\textbf{u}_{i}^\top \theta)}\right|\right\}.
\end{equation}
The LPRE objective function is defined as
\begin{equation}\label{LPRE}
LPRE_n(\theta)=\sum_{n=1}^n \left\{\left|\frac{t_i-\exp(\textbf{u}_i^\top \theta)}{t_i}\right| \times\left|\frac{t_i-\exp(\textbf{u}_{i}^\top \theta)}{\exp(\textbf{u}_{i}^\top \theta)}\right|\right\}.
\end{equation}

Now consider the realistic case with right censoring. For subject $i(=1,\ldots, n)$, let $c_i$ be the censoring variable which is independent of $\textbf{x}_i, \textbf{z}_i$, and $t_i$. We observe $y_i = \min(t_i,c_i)$ and $\delta_i = 1(t_i \le c_i)$. Without loss of generality, assume that the data $(y_i,\delta_i,\textbf{u}_i)$ have been sorted according to $y_i$ from the smallest to the largest.

\subsection{Penalized estimation and selection}
Consider the general relative error (GRE) criterion
\begin{equation}\label{GRE}
GRE_n(\theta)=\sum_{n=1}^n g\left\{\left|\frac{t_i-\exp(\textbf{u}_i^\top \theta)}{t_i}\right|, \left|\frac{t_i-\exp(\textbf{u}_{i}^\top \theta)}{\exp(\textbf{u}_{i}^\top \theta)}\right|\right\},
\end{equation}
where $g(a,b)$ is a bivariate function satisfying certain regularity conditions. When $g(a,b)=a+b$, the GRE criterion becomes the LARE \cite{chen2010least}; when $g(a,b)=ab$, it becomes the LPRE \cite{chen2013least}. 

To accommodate right censoring in estimation, we adopt a weighted approach. Specifically, we first compute the Kaplan-Meier weights $\{w_i\}_{i=1}^n$ as
\begin{equation}
w_1=\frac{\delta_1}{n},w_i=\frac{\delta_i}{n-i+1}\prod_{j=1}^{i-1}
\Big(\frac{n-j}{n-j+1}\Big)^{\delta_j},i=2,\cdots,n.
\end{equation}
We propose the weighted objective function
\begin{equation}
Q_n(\theta)=\sum_{i=1}^n w_i g\left\{\left|\frac{y_i-\exp(\textbf{u}_i^\top \theta)}{y_i}\right|, \left|\frac{y_i-\exp(\textbf{u}_i^\top \theta)}{\exp(\textbf{u}_i^\top \theta)}\right| \right\}.
\end{equation}

In genetic interaction analysis, the dimension of unknown parameters can be much larger than the sample size. For regularized estimation and identification of important effects, we adopt penalization, where the objective function is
\begin{equation}
L_{n,\lambda}(\theta) = Q_n(\theta)+\varphi_\lambda(\theta).
\end{equation}
Here $\varphi_\lambda(\theta)$ is the penalty function. Adopting penalization for genetic interaction analysis has been pursued in recent literature. See for example \cite{bien2013lasso, liu2013identification, shi2014penalized}.

Multiple penalties are potentially applicable. Here we adopt the MCP \cite{zhang2010nearly}, which has been the choice of many high-dimensional studies including genetic interaction analysis. The penalty is defined as $\varphi_\lambda(t)=\lambda\int^{|t|}_0(1-x/(\gamma\lambda))_+dx$. $\gamma>0$ is the regularization parameter, and $\lambda$ is the tuning parameter.

It is noted that applying the MCP may lead to results not respecting the ``main effects, interactions" hierarchy, which has been stressed in some recent studies \cite{bien2013lasso}. The hierarchy postulates that the main effects corresponding to the identified interactions should be automatically identified. This can be achieved by replacing the MCP with for example sparse group penalties. However, we note that the computational cost of such penalties can be much higher. In addition, some published studies have demonstrated pure interactions without the presence of main effects \cite{caspi2006gene,zimmermann2011interaction}. In data analysis, when it is necessary to reinforce the hierarchy, we can refit and add back the main effects corresponding to the identified interactions (if these main effects are not identified in the first place).

\subsection{Computation}

For optimizing the penalized objective function, we propose combining the majorize-minimization (MM) algorithm \cite{hunter2005variable} with the coordinate descent (CD) algorithm \cite{wu2008coordinate}. The MM is used to approximate the objective function using its quadratic majorizer, while the CD is used for iteratively updating the estimate.

Specifically, when $g(a,b)=ab$,  it is easy to compute the gradient and hessian matrix for $Q_n(\theta)$, and so approximation may not be needed. However when $g(a,b)=a+b$, computing the hessian matrix becomes difficult. With the estimate $\theta^{(s)}$ at the beginning of the $s+1$th iteration, we approximate $Q_n(\theta)$ by
\begin{eqnarray*}
Q_n(\theta; \theta^{(s)}) &=& \frac{1}{2}\sum_{i=1}^n w_i\left\{\frac{(1-y_i^{-1}\exp(\textbf{u}_i^\top \theta))^2}{|1-y_i^{-1}\exp(\textbf{u}_i^\top \theta^{(s)})|}+|1-y_i^{-1}\exp(\textbf{u}_i^\top \theta^{(s)})|\right.\\
&& \left.+ \frac{({1-y_i\exp(-\textbf{u}_i^\top \theta)})^2}{|{1-y_i\exp(-\textbf{u}_i^\top \theta^{(s)})}|} + |{1-y_i\exp(-\textbf{u}_i^\top \theta^{(s)})}| \right\}.
\end{eqnarray*}
It can be shown that $Q_n(\theta; \theta^{(s)})\geq Q_n(\theta)$, and the equality holds if and only if $\theta^{(s)}= \theta$. For the MCP, we use a quadratic approximation
\[
\varphi_{\lambda}(\theta; \theta^{(s)}) =  \varphi_{\lambda}(\theta^{(s)})+ \frac{1}{2 |\theta^{(s)}|}\varphi'_{\lambda}(\theta^{(s)})(\theta^2-\theta^{(s)2}).
\]
By ignoring terms  not related to $\theta$ in $Q_n(\theta; \theta^{(s)})+ \varphi_{\lambda}(\theta; \theta^{(s)})$, we have a smooth loss function $L_{n, \lambda}(\theta; \theta^{(s)})$, which is
\begin{equation}\label{app}
\sum_{i=1}^n w_i\left\{\frac{(1-y_i^{-1}\exp(\textbf{u}_i^\top \theta))^2}{|1-y_i^{-1}\exp(\textbf{u}_i^\top \theta^{(s)})|}+ \frac{({1-y_i\exp(-\textbf{u}_i^\top \theta)})^2}{|{1-y_i\exp(-\textbf{u}_i^\top \theta^{(s)})}|}  \right\} + \frac{1}{ |\theta^{(s)}|}\varphi'_{\lambda}(\theta^{(s)})\theta^2.
\end{equation}
To solve the minimization problem $\theta^{(s+1)}= \arg \min_{\theta}L_{n, \lambda}(\theta; \theta^{(s)})$, we employ the coordinate descent algorithm. In summary, the algorithm proceeds as follows:

\medskip\noindent
Step 1. Initialize $s=0$. Compute $\theta^{(0)}$ as the Lasso estimate (which can be viewed as an extreme case of the MCP estimate).

\noindent
Step 2. Apply the CD algorithm to minimize the loss function $L_{n, \lambda}(\theta; \theta^{(s)})$ in (\ref{app}). Denote the estimate as $\theta^{(s+1)}$. Specially, the CD algorithm updates one coordinate at a time and treats the other coordinates as fixed. Define $u_{ij}$ as the $j$th component of $\textbf{u}_i$. For $j \in \{1, \cdots, p+q+pq\}$, defined $\vartheta_{i,-j}=\sum_{t<j}u_{it} \theta_{t}^{(s+1)} + \sum_{t>j}u_{it} \theta_{t}^{(s)}$, then
\begin{eqnarray*}
  \theta_j^{(s+1)} &=& \arg\min_{\theta_j} \left\{\sum_{i=1}^n w_i \left[\frac{(1-y_i^{-1}\exp(\vartheta_{i,-j} +u_{ij}\theta_j ))^2}{|1-y_i^{-1}\exp(\textbf{u}_i^\top \theta^{(s)})|} \right.\right.\\
  && \left.\left.~~~~~~+ \frac{({1-y_i\exp(-\vartheta_{i,-j}-u_{ij}\theta_j )})^2}{|{1-y_i\exp(-\textbf{u}_i^\top \theta^{(s)})}|} \right] + \frac{1}{ |\theta_j^{(s)}|}\varphi'_{\lambda}(\theta_j^{(s)})\theta_j^2\right\} ~.
\end{eqnarray*}

\noindent
Step 3. Repeat Step 2 until convergence. We use the $L_2$-norm of the difference between two consecutive estimates less than $10^{-6}$ as the convergence criterion.

The proposed method involves tunings. For $\gamma$, published studies \cite{zhang2010nearly} suggest selecting from a small number of values or fixing it. In our simulation, we find that the estimation results are not sensitive to the value of $\gamma$. We follow published studies and set $\gamma=6$. The selection of $\lambda$ will be described in the following sections.

\section{Simulation}
Beyond evaluating performance of the proposed approach, we also use simulation to compare with the penalized weighted least squares (simply denoted as LS) and penalized weighted least absolute deviation (denoted as LAD) methods, which respectively have objective functions
\[\sum_{i=1}^n w_i(\log(y_i)-\textbf{u}_i^\top\theta)^2+\varphi_\lambda(\theta)
~~\mbox{and}
~~
\sum_{i=1}^n w_i|\log(y_i)-\textbf{u}_i^\top\theta|+\varphi_\lambda(\theta),\]
where $\{w_i\}_{i=1}^n$ and $\varphi_\lambda(\theta)$ are the same as defined before. \\

\noindent{\bf Simulation I.} In model
$t_i=\exp(\textbf{x}_i^\top\alpha+\textbf{z}_i^\top\beta+(\textbf{x}_i\otimes \textbf{z}_i)^\top\xi)\varepsilon_i,~ i=1,\cdots,n,$ $\textbf{z}_i$'s have a multivariate normal distribution with marginal means 0 and marginal variances 1. Denote the correlation coefficient between genes $j$ and $k$ as $\rho_{jk}$. Consider the following correlation structures: (i) independent, where $\rho_{jk}=0$ if $j\ne k$, (ii) AR(0.2), where $\rho_{jk}=0.2^{|j-k|}$; (iii) AR(0.8), where $\rho_{jk}=0.8^{|j-k|}$; (iv) Band1, where $\rho_{jk}=0.3$ if $|j-k|=1$ and $\rho_{jk}=0$ otherwise; and (v) Band2, where $\rho_{jk}=0.6$ if $|j-k|=1$, $\rho_{jk}=0.3$ if $|j-k|=2$, and $\rho_{jk}=0$ otherwise. We generate $\textbf{x}_i$'s from the standard multivariate normal distribution. We set $n=200$, $q=5$, and $p=500$. The dimension of genetic effects and interactions is much larger than the sample size. There are a total of 35 nonzero effects: 5 main effects of the E factors, 10 main effects of the G factors, and 20 interactions. The nonzero coefficients are randomly generated from $Uniform(0.4,1.2)$. We consider two error distributions: (i) $\log(\varepsilon)$ follows $N(0,1)$, and (ii) $\log(\varepsilon)$ follows $Unif(-2,2)$. The event times are computed from the AFT model. The censoring times are generated from a uniform distribution, with a censoring rate about 20\%.

\medskip\noindent{\bf Simulation II.} Data are first generated in the same manner as under Simulation I. To mimic discrete genetic data (for example SNPs), we dichotomize the simulated genetic data at -1 and 0.5 to create three levels.

\begin{table}[h]
\caption{Summary of Simulation II. In each cell, mean (sd) based on 200 replicates}
\vskip .1in
\label{table3}
\centering 
\setlength{\tabcolsep}{3pt}
\begin{tabular}{c c c c c c}\hline
          &&AUC&SE&TPR&FPR\\ \hline
\multicolumn{6}{c}{$\log(\varepsilon)\sim N(0,1)$} \\
\hline
independent&LARE&0.846(0.031)&19.53(3.321)&0.601(0.063)&0.098(0.013)\\
&LPRE&0.837(0.032)&19.47(3.130)&0.572(0.171)&0.095(0.134)\\
&LAD&0.833(0.029)&20.26(3.118)&0.564(0.117)&0.084(0.026)\\
&LS&0.854(0.020)&20.78(2.641)&0.562(0.109)&0.076(0.013)\\\hline
AR(0.2)&LARE&0.868(0.034)&17.55(3.252)&0.739(0.082)&0.103(0.018)\\
&LPRE&0.863(0.024)&16.68(3.671)&0.649(0.153)&0.062(0.027)\\
&LAD&0.847(0.027)&19.57(2.947)&0.564(0.100)&0.078(0.026)\\
&LS&0.860(0.024)&18.66(2.583)&0.628(0.086)&0.071(0.011)\\\hline
AR(0.8)&LARE&0.928(0.029)&7.655(2.611)&0.891(0.053)&0.062(0.027)\\
&LPRE&0.898(0.032)&7.755(2.990)&0.871(0.076)&0.066(0.021)\\
&LAD&0.911(0.022)&13.68(2.973)&0.758(0.098)&0.069(0.023)\\
&LS&0.901(0.026)&12.74(2.417)&0.779(0.104)&0.063(0.019)\\\hline
Band1&LARE&0.868(0.033)&18.51(3.316)&0.673(0.080)&0.078(0.022)\\
&LPRE&0.859(0.026)&17.78(3.560)&0.641(0.143)&0.059(0.023)\\
&LAD&0.850(0.031)&19.27(3.676)&0.629(0.119)&0.085(0.025)\\
&LS&0.864(0.022)&18.92(2.853)&0.616(0.074)&0.078(0.012)\\\hline
Band2&LARE&0.904(0.028)&10.82(2.571)&0.828(0.158)&0.060(0.017)\\
&LPRE&0.875(0.031)&11.39(2.922)&0.787(0.102)&0.055(0.021)\\
&LAD&0.872(0.033)&17.68(3.673)&0.685(0.108)&0.075(0.027)\\
&LS&0.880(0.025)&16.92(3.114)&0.725(0.081)&0.075(0.014)\\
\hline
\multicolumn{6}{c}{$\log(\varepsilon)\sim Unif(-2,2)$} \\
\hline
independent&LARE&0.840(0.032)&19.38(3.024)&0.634(0.073)&0.111(0.024)\\
&LPRE&0.845(0.022)&20.46(2.898)&0.582(0.169)&0.094(0.035)\\
&LAD&0.831(0.033)&21.29(3.453)&0.569(0.123)&0.081(0.027)\\
&LS&0.847(0.021)&21.03(3.258)&0.557(0.087)&0.080(0.018)\\\hline
AR(0.2)&LARE&0.832(0.029)&18.63(3.286)&0.696(0.076)&0.093(0.019)\\
&LPRE&0.850(0.022)&18.15(4.075)&0.616(0.082)&0.083(0.012)\\
&LAD&0.835(0.028)&19.49(2.958)&0.583(0.127)&0.082(0.028)\\
&LS&0.858(0.021)&20.52(3.063)&0.587(0.111)&0.076(0.018)\\\hline
AR(0.8)&LARE&0.913(0.031)&9.610(2.219)&0.833(0.128)&0.068(0.023)\\
&LPRE&0.889(0.025)&8.732(2.770)&0.857(0.105)&0.052(0.016)\\
&LAD&0.900(0.030)&15.85(2.980)&0.736(0.124)&0.072(0.026)\\
&LS&0.895(0.026)&14.60(2.970)&0.732(0.108)&0.007(0.029)\\\hline
Band1&LARE&0.850(0.028)&14.23(3.010)&0.714(0.082)&0.097(0.020)\\
&LPRE&0.856(0.023)&15.64(3.274)&0.624(0.120)&0.083(0.016)\\
&LAD&0.844(0.030)&20.94(3.371)&0.543(0.114)&0.077(0.024)\\
&LS&0.856(0.023)&19.70(2.899)&0.626(0.090)&0.076(0.011)\\\hline
Band2&LARE&0.868(0.032)&13.06(3.513)&0.782(0.163)&0.098(0.025)\\
&LPRE&0.864(0.030)&12.23(3.713)&0.763(0.148)&0.057(0.024)\\
&LAD&0.870(0.029)&17.23(3.555)&0.680(0.128)&0.073(0.027)\\
&LS&0.869(0.033)&16.46(2.470)&0.704(0.093)&0.071(0.010)\\
\hline
\end{tabular}
\end{table}

We evaluate the simulation results in two ways. First, we consider a sequence of $\lambda$ values, evaluate identification performance at each value, and then compute the overall AUC (area under the ROC -- receiver operating characteristic -- curve). In addition, we also select the optimal $\lambda$ using a cross validation approach and then compute the estimation squared error (SE), true positive rate (TPR), and false positive rate (TPR) at the optimal tuning. The summary based on 200 replicates is provided in Table 1 and 3 (Appendix), respectively. Simulation suggests that, when evaluated using AUC, the four methods have similar performance. Under Simulation I, the performance is also similar in terms of SE, TPR, and FPR. However, under Simulation II, the proposed LARE and LPRE can have better performance. In addition, it is also observed that LARE may outperform LPRE, at the cost of slightly higher computer time. Overall simulation suggests that the proposed approach, especially LARE, performs comparable to or better than the alternatives. Thus it provides a ``safe" choice for practical data analysis.

\section{Analysis of lung cancer prognosis data}

Lung cancer is the leading cause of cancer death worldwide. Genetic profiling studies have been extensively conducting, searching for genetic risk factors associated with lung cancer prognosis. Here we analyze the TCGA (The Cancer Genome Atlas) data on the prognosis of lung adenocarcinoma. The TCGA data were recently collected and published by NCI and have a high quality. The prognosis outcome of interest is overall survival. The dataset contains records on 468 patients, among whom 117 died during follow-up. The median follow-up time is 8 months.

\begin{table}[h]
\caption{Analysis of lung cancer data with LARE: main genetic effects and G$\times$E interactions. For the interactions, values in ``()'' are the stability results.}
\vskip .1in
\label{table5}
\centering 
\setlength{\tabcolsep}{3pt}
\begin{tabular}{c c c c c c }\hline

          &&\multicolumn{4}{c}{Interactions}\\
          \cline{3-6}
Gene &Main effects&Age&Gender&Smoking pack year&Smoking history\\\hline
ADORA2B&	-0.231	&&&&	-0.260(0.76)\\
AKIRIN2&	-0.281	\\
ASB12&	-0.241	\\
C5ORF45&	-0.042	\\
C14ORF93&	-0.472\\
C16ORF93&	-0.160&	-0.293(0.91)	\\
CAND1&	0.309	&-2.181(0.95)	\\
CBWD2&	0.234	\\
CDR2&	0.210	\\
CIAPIN1&	0.187&&	&		-0.179(0.85)	\\
DCP1B&	0.448	\\
DYRK2&	-1.41&	0.758(0.66)	\\
EIF4EBP1&	0.081&	-0.001(0.81)\\
EMB&	0.224	\\
FDXR&	0.293	&&&	-0.477(0.99)	\\
GALK2&	-0.158	&&&	-0.240(0.75)	\\
GOLGA7&	-0.146	&-0.096(0.45)	\\
HERPUD2&	0.121	\\
HOXC13&	-0.248&	-0.145(0.98)	\\
ING1&	-2.117	&&&&		2.154(0.97)\\
INO80B&	-0.164	&&&	-1.607(0.95)	\\
KIF21B&	-0.391	&-0.446(0.99)	\\
KLHDC1&	-0.011	&&&	0.382(0.98)\\
LIG4&	-0.584	&&	0.299(0.80)	\\
LINC00471&	0.236	&&&&	0.114(0.94)\\
LINC00476&	0.258	&&&&	0.056(0.55)\\
LRRC45&	-0.136	&-0.083(0.93)	\\
MCAT&	0.103	&&&	0.180(0.96)	\\
MVD&	-0.348	\\
NCALD&	0.376	&&	-0.605(0.70)	\\
OTUD1&	0.189	&&&	0.038(0.34)	\\
PEX19&	-0.444	&&&&	0.045(0.55)\\
PHLPP1&	-0.439	\\
PNPLA2&	-0.193	&0.014(0.55)	\\
PPM1A&	-0.124	&&&	0.166(0.89)	\\
PPP2R2D&	0.157&	-0.234(0.67)	\\
RBM11&	0.032	&&	-0.291(0.71)	\\
RNF6&	-0.215	&0.199(0.90)	\\
RNF126P1&	0.225	\\
RPS27&	0.134	&&&&	-0.155(0.22)\\
SCAND2P&	-0.002	&&&	0.329(0.35)	\\
SERTAD4&	-0.356	&&	0.350(0.91)	\\
SGSM3&	0.285	&&&	-0.039(0.46)\\
SH3RF1&	-0.096	\\
SLC25A2&	-0.009	&&	-0.335(0.94)	\\
SPCS3&	-0.310	&&&&	0.340(0.66)\\
SPRED2&	-0.260	\\
SRRM3&	-0.317	&&&&	-0.244(0.70)\\
TXN2&	-0.339	&&&	0.012(0.46)	\\
UBE4B&	0.418	&-0.497(0.53)	\\
VPS13B&	0.065	&&&	-0.108(0.99)	\\
ZNF727&	0.401&	-0.254(0.78)	\\
\hline
\end{tabular}
\end{table}

Four E factors are included in analysis: age, gender, smoking pack years, and smoking history. All four have been suggested as associated with lung cancer prognosis in the literature. Among them, age and smoking pack years are continuous and normalized prior to analysis. Gender and smoking history are binary. A total of 436 subjects have complete E measurements. Among them, 110 died during follow-up, and the median follow-up time is 23 months. For the 326 censored subjects, the median follow-up time is 6 months.

Measurements on 18,897 gene expressions are available. To improve stability and reduce computational cost, we conduct marginal prescreening based on genes' univariate regression significance (p-value less than or equal to 0.1) and interquartile range (above the median of all interquartile ranges). Similar procedures have been adopted in the literature. A total of 819 gene expressions are included in downstream analysis. For each gene expression, we normalize to have mean 0 and variance 1.

We apply the proposed approach and select the optimal $\lambda$ using five-fold cross validation. The detailed identification and estimation results are presented in Tables 2 (LARE) and 5 (LPRE, Appendix). As previously described, it is possible that the main effects corresponding to the identified interactions are not identified. To respect the ``main effects, interactions" hierarchy, we add back such main effects and re-fit. Beyond the proposed, we also apply the LS and LAD methods. The summary of applying different methods is provided in Table \ref{table9} (Appendix). Detailed estimation and identification results using the alternatives are presented in Tables \ref{table7} and \ref{table8} (Appendix). Different methods identify different sets of main effects and interactions. It is interesting that all of the main effects and interactions identified by LPRE are identified by LARE. They may represent more reliable findings. The LAD method identifies much fewer effects.

To complement the identification and estimation analysis, we evaluate stability. Specifically, we randomly remove ten subjects and then analyze data. This procedure is repeated 200 times. We then compute the probability that an interaction term is identified. Such an evaluation has been conducted in the literature. The stability results are provided in Tables 2 and 5-7(Appendix). We can see that most of the identified interactions are relatively stable, with many having probabilities of being identified close to one.

\section{Discussion}

The identification of important G$\times$E interactions remains a challenging problem. In this article, we have introduced using the relative error criteria as loss functions. A penalized approach has been adopted for estimation and selection. Simulation shows that the proposed approach has performance comparable to or better than the alternatives. Thus it may be provide a useful alternative for data analysis. A limitation of this study is that the asymptotic properties have not been established. In the analysis of a lung cancer dataset, the LARE and LPRE results are relatively consistent but different from the alternatives. The identified interactions are reasonably stable. More examination of the findings is needed in the future.

\begin{acknowledgement}
We thank the participants of Joint 24th ICSA Applied Statistics Symposium and 13th Graybill Conference in Colorado and organizers of this proceedings. This study was supported in part by the National Science Foundation of China (Grant No. 11401561), National Social Science Foundation of China (13CTJ001, 13\&ZD148), NIH (CA016359, CA191383), and the U.S. VA Cooperative Studies Program of the Department of Veterans Affairs, Office of Research and Development.
\end{acknowledgement}

\bibliographystyle{spmpsci}
\bibliography{mybib}

\begin{thebibliography}{10}
\providecommand{\url}[1]{{#1}}
\providecommand{\urlprefix}{URL }
\expandafter\ifx\csname urlstyle\endcsname\relax
  \providecommand{\doi}[1]{DOI~\discretionary{}{}{}#1}\else
  \providecommand{\doi}{DOI~\discretionary{}{}{}\begingroup
  \urlstyle{rm}\Url}\fi

\bibitem{bien2013lasso}
Bien, J., Taylor, J., Tibshirani, R., et~al.: A lasso for hierarchical
  interactions.
\newblock The Annals of Statistics \textbf{41}(3), 1111--1141 (2013)

\bibitem{caspi2006gene}
Caspi, A., Moffitt, T.E.: Gene--environment interactions in psychiatry: joining
  forces with neuroscience.
\newblock Nature Reviews Neuroscience \textbf{7}(7), 583--590 (2006)

\bibitem{chen2010least}
Chen, K., Guo, S., Lin, Y., Ying, Z.: Least absolute relative error estimation.
\newblock Journal of the American Statistical Association \textbf{105}(491),
  1104--1112 (2010)

\bibitem{chen2013least}
Chen, K., Lin, Y., Wang, Z., Ying, Z.: Least product relative error estimation.
\newblock arXiv preprint arXiv:1309.0220  (2013)

\bibitem{cordell2009detecting}
Cordell, H.J.: Detecting gene--gene interactions that underlie human diseases.
\newblock Nature Reviews Genetics \textbf{10}(6), 392--404 (2009)

\bibitem{hunter2005gene}
Hunter, D.J.: Gene--environment interactions in human diseases.
\newblock Nature Reviews Genetics \textbf{6}(4), 287--298 (2005)

\bibitem{hunter2005variable}
Hunter, D.R., Li, R.: Variable selection using mm algorithms.
\newblock Annals of statistics \textbf{33}(4), 1617--1642 (2005)

\bibitem{khoshgoftaar1992predicting}
Khoshgoftaar, T.M., Bhattacharyya, B.B., Richardson, G.D.: Predicting software
  errors, during development, using nonlinear regression models: a comparative
  study.
\newblock Reliability, IEEE Transactions on \textbf{41}(3), 390--395 (1992)

\bibitem{li2014empirical}
Li, Z., Lin, Y., Zhou, G., Zhou, W.: Empirical likelihood for least absolute
  relative error regression.
\newblock Test \textbf{23}(1), 86--99 (2014)

\bibitem{liu2013identification}
Liu, J., Huang, J., Zhang, Y., Lan, Q., Rothman, N., Zheng, T., Ma, S.:
  Identification of gene--environment interactions in cancer studies using
  penalization.
\newblock Genomics \textbf{102}(4), 189--194 (2013)

\bibitem{north2008importance}
North, K.E., Martin, L.J.: The importance of gene¡ªenvironment interaction
  implications for social scientists.
\newblock Sociological Methods \& Research \textbf{37}(2), 164--200 (2008)

\bibitem{park1998relative}
Park, H., Stefanski, L.: Relative-error prediction.
\newblock Statistics \& probability letters \textbf{40}(3), 227--236 (1998)

\bibitem{shi2014penalized}
Shi, X., Liu, J., Huang, J., Zhou, Y., Xie, Y., Ma, S.: A penalized robust
  method for identifying gene--environment interactions.
\newblock Genetic epidemiology \textbf{38}(3), 220--230 (2014)

\bibitem{thomas2010methods}
Thomas, D.: Methods for investigating gene-environment interactions in
  candidate pathway and genome-wide association studies.
\newblock Annual review of public health \textbf{31}, 21--36 (2010)

\bibitem{tsionas2014bayesian}
Tsionas, E.G.: Bayesian analysis of least absolute relative error regression.
\newblock Communications in Statistics-Theory and Methods \textbf{43}(23),
  4988--4997 (2014)

\bibitem{van2015relative}
Van~Dam, L.C., Ernst, M.O.: Relative errors can cue absolute visuomotor
  mappings.
\newblock Experimental brain research \textbf{233}(12), 3367--3377 (2015)

\bibitem{wu2014integrative}
Wu, C., Cui, Y., Ma, S.: Integrative analysis of gene--environment interactions
  under a multi-response partially linear varying coefficient model.
\newblock Statistics in medicine \textbf{33}(28), 4988--4998 (2014)

\bibitem{wu2014selective}
Wu, C., Ma, S.: A selective review of robust variable selection with
  applications in bioinformatics.
\newblock Briefings in bioinformatics \textbf{16}, 873--883 (2015)

\bibitem{wu2008coordinate}
Wu, T.T., Lange, K.: Coordinate descent algorithms for lasso penalized
  regression.
\newblock The Annals of Applied Statistics pp. 224--244 (2008)

\bibitem{xia2015regularized}
Xia, X., Liu, Z., Yang, H.: Regularized estimation for the least absolute
  relative error models with a diverging number of covariates.
\newblock Computational Statistics \& Data Analysis  (2015)

\bibitem{zhang2010nearly}
Zhang, C.: Nearly unbiased variable selection under minimax concave penalty.
\newblock The Annals of Statistics pp. 894--942 (2010)

\bibitem{zhang2013local}
Zhang, Q., Wang, Q.: Local least absolute relative error estimating approach
  for partially linear multiplicative model.
\newblock Statistica Sinica \textbf{23}, 1091--1116 (2013)

\bibitem{zhu2014identifying}
Zhu, R., Zhao, H., Ma, S.: Identifying gene--environment and gene--gene
  interactions using a progressive penalization approach.
\newblock Genetic epidemiology \textbf{38}(4), 353--368 (2014)

\bibitem{zimmermann2011interaction}
Zimmermann, P., Br{\"u}ckl, T., Nocon, A., Pfister, H., Binder, E.B., Uhr, M.,
  Lieb, R., Moffitt, T.E., Caspi, A., Holsboer, F., et~al.: Interaction of
  fkbp5 gene variants and adverse life events in predicting depression onset:
  results from a 10-year prospective community study.
\newblock American Journal of Psychiatry \textbf{168}, 1107--1116 (2011)

\end{thebibliography}

\clearpage
\section*{Appendix}

\begin{table}[h]
\caption{Summary of Simulation I. In each cell, mean (sd) based on 200 replicates.}
\vskip .1in
\label{table1}
\centering 
\setlength{\tabcolsep}{3pt}
\begin{tabular}{c c c c c c}\hline

          &&AUC&SE&TPR&FPR\\ \hline
\multicolumn{6}{c}{$\log(\varepsilon)\sim N(0,1)$} \\
\hline

independent&LARE&0.835(0.033)&10.58(1.742)&0.639(0.085)&0.092(0.018)\\
&LPRE&0.837(0.032)&11.64(1.918)&0.603(0.077)&0.080(0.011)\\
&LAD&0.848(0.033)&10.63(1.832)&0.599(0.089)&0.076(0.019)\\
&LS&0.836(0.031)&10.15(1.872)&0.590(0.089)&0.079(0.019)\\\hline
AR(0.2)&LARE&0.859(0.038)&9.522(2.475)&0.697(0.135)&0.071(0.024)\\
&LPRE&0.858(0.036)&11.13(2.080)&0.660(0.126)&0.064(0.024)\\
&LAD&0.877(0.033)&9.363(2.230)&0.672(0.109)&0.079(0.015)\\
&LS&0.858(0.033)&8.972(1.929)&0.661(0.112)&0.066(0.018)\\\hline
AR(0.8)&LARE&0.920(0.032)&5.834(2.577)&0.844(0.161)&0.057(0.028)\\
&LPRE&0.922(0.032)&6.932(2.116)&0.833(0.122)&0.039(0.028)\\
&LAD&0.939(0.025)&5.268(1.744)&0.835(0.127)&0.051(0.027)\\
&LS&0.923(0.028)&5.598(1.744)&0.795(0.127)&0.036(0.029)\\\hline
Band1&LARE&0.860(0.033)&9.793(2.599)&0.721(0.136)&0.088(0.020)\\
&LPRE&0.860(0.033)&9.338(1.871)&0.698(0.118)&0.064(0.019)\\
&LAD&0.883(0.028)&8.455(1.623)&0.690(0.111)&0.072(0.016)\\
&LS&0.862(0.031)&8.573(2.849)&0.674(0.141)&0.060(0.023)\\\hline
Band2&LARE&0.904(0.028)&6.907(2.262)&0.784(0.172)&0.072(0.021)\\
&LPRE&0.893(0.033)&6.706(2.842)&0.741(0.138)&0.052(0.021)\\
&LAD&0.915(0.033)&6.638(2.131)&0.757(0.142)&0.058(0.022)\\
&LS&0.899(0.034)&6.984(2.112)&0.746(0.147)&0.049(0.028)\\
\hline
\multicolumn{6}{c}{$\log(\varepsilon)\sim Unif(-2,2)$} \\
\hline
independent&LARE&0.830(0.034)&12.13(2.716)&0.647(0.092)&0.082(0.016)\\
&LPRE&0.841(0.036)&10.64(2.592)&0.597(0.118)&0.070(0.021)\\
&LAD&0.849(0.028)&10.74(1.886)&0.570(0.136)&0.079(0.025)\\
&LS&0.835(0.029)&11.15(2.094)&0.540(0.135)&0.069(0.027)\\\hline
AR(0.2)&LARE&0.846(0.027)&9.231(1.574)&0.657(0.139)&0.088(0.018)\\
&LPRE&0.854(0.031)&10.91(2.148)&0.628(0.120)&0.076(0.025)\\
&LAD&0.872(0.031)&9.846(1.416)&0.628(0.127)&0.072(0.024)\\
&LS&0.852(0.034)&9.721(1.772)&0.599(0.153)&0.062(0.026)\\\hline
AR(0.8)&LARE&0.923(0.030)&6.454(1.449)&0.793(0.150)&0.048(0.026)\\
&LPRE&0.921(0.029)&7.832(1.893)&0.792(0.180)&0.035(0.032)\\
&LAD&0.934(0.027)&6.007(1.452)&0.798(0.123)&0.054(0.022)\\
&LS&0.917(0.027)&6.932(2.101)&0.772(0.207)&0.034(0.030)\\\hline
Band1&LARE&0.851(0.033)&9.944(1.521)&0.683(0.117)&0.079(0.028)\\
&LPRE&0.847(0.033)&9.832(1.931)&0.658(0.108)&0.075(0.039)\\
&LAD&0.878(0.029)&9.126(1.865)&0.694(0.117)&0.080(0.020)\\
&LS&0.857(0.033)&9.465(1.816)&0.662(0.121)&0.091(0.032)\\\hline
Band2&LARE&0.877(0.034)&6.803(1.303)&0.797(0.171)&0.070(0.023)\\
&LPRE&0.889(0.032)&6.943(1.503)&0.763(0.129)&0.058(0.023)\\
&LAD&0.911(0.032)&6.439(1.458)&0.760(0.148)&0.062(0.021)\\
&LS&0.893(0.027)&6.498(1.860)&0.733(0.162)&0.057(0.027)\\\hline
\end{tabular}
\end{table}

\clearpage
\begin{table}[h]
\caption{Analysis of lung cancer data using different methods: the numbers of identified main effects and interactions and overlaps. In each cell, number of identified main effects/number of identified interactions.}
\vskip .1in
\label{table9}
\centering 
\setlength{\tabcolsep}{3pt}
\begin{tabular}{c c c c c}\hline

          &LARE & LPRE & LAD & LS\\\hline
LARE&52/38&43/32&8/3&23/12\\
LPRE&&43/32&7/3&20/10\\
LAD&&&34/13&14/6\\
LS&&&&45/32\\
\hline
\end{tabular}
\end{table}

\clearpage
\begin{table}[h]
\caption{Analysis of lung cancer data with LPRE: main genetic effects and G$\times$E interactions. For the interactions, values in ``()'' are the stability results.}
\vskip .1in
\label{table6}
\centering 
\setlength{\tabcolsep}{3pt}
\begin{tabular}{c c c c c c }\hline

          &&\multicolumn{4}{c}{Interactions}\\
          \cline{3-6}
Probe &Main effects&Age&Gender&Smoking pack year&Smoking history\\\hline
ADORA2B&	0.548&&&&-1.093(0.73)\\
AKIRIN2&	-0.287	\\
ASB12&	-0.156	\\
C14ORF93&	-0.347	\\
C16ORF93&	-0.023	&-0.290(0.85)	\\
CAND1&	0.536&	-2.113(0.98)	\\
CBWD2&	0.293	\\
CDR2&	0.200	\\
CIAPIN1&	0.163&&&-0.369(0.88)	\\
DCP1B&	0.482	\\
DYRK2&	-1.473	&0.523(0.57)	\\
FDXR&	0.236	&&&	-0.655(0.98)	\\
GALK2&	-0.021&&&	-0.213(0.58)	\\
GOLGA7&	0.037	&-0.091(0.41)	\\
HERPUD2&	0.065	\\
HOXC13&	-0.064	&-0.173(0.98)	\\
ING1&	-1.133	&&&&	0.939(0.96)\\
INO80B&	-0.100	&&&	-0.803(0.75)	\\
KIF21B&	-0.176	&-0.330(0.98)	\\
KLHDC1&	0.030&&&	0.232(0.81)	\\
LIG4&	-0.446	&&	0.088(0.96)	\\
LINC00471&	0.037	&&&&	0.146(0.71)\\
LRRC45&	-0.249	&-0.234(0.86)	\\
MCAT&	0.017	&&&	0.082(0.65)	\\
MVD&	-0.321	\\
NCALD&	0.057	&&	-0.436(0.61)	\\
OTUD1&	0.131	&&&	0.094(0.32)	\\
PEX19&	-0.535	&&&&	0.128(0.36)\\
PHLPP1&	-0.355	\\
PNPLA2&	-0.073&	0.009(0.18)	\\
PPP2R2D&	0.137&	-0.173(0.62)	\\
RBM11&	0.119	&&	-0.263(0.46)	\\
RNF6&	-0.256	&0.303(0.85)	\\
SERTAD4&	-0.040&&	0.148(0.60)	\\
SGSM3&	0.176	&&&	-0.004(0.16)	\\
SH3RF1&	-0.177\\
SLC25A2&	0.040	&&	-0.314(0.78)	\\
SPCS3&	-0.559	&&&&	0.351(0.42)\\
SPRED2&	-0.437	\\
SRRM3&	0.337	&&&&	-0.744(0.80)\\
TXN2&	-0.225	&&&	0.160(0.32)	\\
UBE4B&	0.327	&-0.524(0.50)	\\
VPS13B&	0.188	&&&	-0.406(0.91)	\\
\hline
\end{tabular}
\end{table}

\clearpage
\begin{table}[h]
\caption{Analysis of lung cancer data with LAD: main genetic effects and G$\times$E interactions. For the interactions, values in ``()'' are the stability results.}
\vskip .1in
\label{table7}
\centering 
\setlength{\tabcolsep}{3pt}
\begin{tabular}{c c c c c c }\hline

          &&\multicolumn{4}{c}{Interactions}\\
          \cline{3-6}
Probe &Main effects&Age&Gender&Smoking pack year&Smoking history\\\hline
ADORA2B&	-0.053	\\
AKR1A1&	-0.085	\\
ALG9&	-0.072	\\
ANKRD54&	0.012	\\
ANP32B&	-0.040	\\
ARFGAP2&	-0.034	\\
ARL15&	0.029	\\
ASB12&	-0.014	\\
ASCC1&	0.003	\\
ATP8B2&	0.023	\\
C2ORF16&	-0.032	&&	-0.032(0.37)	&&	0.029(0.59)\\
C2ORF42&	0.055	\\
VIPAS39&	0.044	&-0.030(0.59)	\\
C16ORF93&	-0.011&	-0.084(0.86)\\
CAND1&	-0.001	&-0.362(1.00)	\\
CD46&	-0.001	\\
CHKA&	-0.041	\\
DCP1B&	0.050	&&&&	0.036(0.87)\\
DNAJC21&	0.035	\\
DPY19L1&	0.030	&-0.026(0.69)	\\
DUSP6&	-0.007	\\
EIF3F&	-0.009	\\
EMB&	-0.157	\\
FCRLB&	-0.050	\\
FDXR&	-0.009	&&&	-0.159(0.96)	\\
GABPA&	-0.095	\\
GINS4&	-0.069	\\
HKR1&	-0.011	&&&&	-0.008(0.13)\\
KLF10&	-0.028	&0.087(0.73)	\\
LIN37&	0.038	&-0.013(0.97)	\\
LINC00515&	0.029&	-0.043(0.97)	\\
PAF1&	-0.053	\\
SPRED2&	-0.164	&&&&	0.046(0.77)\\
TWISTNB&	0.079	\\
\hline
\end{tabular}
\end{table}

\clearpage
\begin{table}[h]
\caption{Analysis of lung cancer data with LS: main genetic effects and G$\times$E interactions. For the interactions, values in ``()'' are the stability results.}
\vskip .1in
\label{table8}
\centering 
\setlength{\tabcolsep}{3pt}
\begin{tabular}{c c c c c c }\hline

          &&\multicolumn{4}{c}{Interactions}\\
          \cline{3-6}
Probe &Main effects&Age&Gender&Smoking pack year&Smoking history\\\hline
ADORA2B&	-0.027	\\
ARL15&	0.012	&&	-0.046(0.69)	\\
C2ORF16&	0.005	&&	-0.013(0.53)	\\
C11ORF52&	0.029	&&	-0.032(0.49)	\\
C14ORF93&	-0.112	\\
C16ORF93&	-0.017&	-0.096(0.99)	\\
CAND1&	-0.035	&-0.347(0.97)	\\
CBWD2&	0.000	&&0.027(0.58)	\\
CCDC171&	0.011	&&&	-0.026(0.92)	\\
CDR2&	-0.003	\\
DCP1B&	0.103	\\
DNAJB13&	0.045	\\
DNAJC30&	-0.024	&&&	0.025(0.85)	\\
DYRK1B&	0.011	&-0.030(0.58)	\\
EIF3F&	-0.008	\\
EIF4EBP1&	-0.007	&-0.025(0.97)	\\
EMB&	-0.081	\\
FDXR&	-0.009	&&&	-0.135(1.00)	\\
GEMIN8&	0.036	&&	-0.045(0.57)	\\
HIST1H2AJ&	-0.002&	-0.013(0.25)	\\
HNRNPDL&	-0.019	&&&&	-0.010(0.51)\\
HOXC13&	-0.006	&-0.003(0.43)	\\
ING1&	0.016	&&	-0.031(0.36)	\\
INO80B&	-0.004	&&&	-0.282(0.96)	\\
KLHDC1&	0.000	&&&	0.042(0.79)	\\
KLHL7&	0.015	\\
LIN37&	0.014	&-0.050(0.63)	\\
LINC00515&	0.028&	-0.037(0.92)	\\
LRRC45&	0.016	&-0.006(0.41)&&	0.024(0.81)	\\
MVD&	-0.050	\\
PAF1&	0.054	&&	-0.125(0.84)	\\
PHLPP1&	-0.034	\\
PIK3CB&	-0.032	&&&&	-0.007(0.76)\\
PNPLA2&	-0.014	\\
POLN&	-0.011	&&&&	0.030(0.64)\\
PPHLN1&	0.047	\\
RNF6&	0.014	&0.007(0.92)	\\
RPS27&	0.027	&&&&	-0.092(0.76)\\
SGSM3&	-0.003	&&&	-0.020(0.37)	\\
SPRED2&	-0.047	\\
SYNCRIP&	-0.042	&0.029(0.28)	\\
TRAM1L1&	0.002	&&	-0.044(0.81)	\\
TWISTNB&	0.031	\\
UBE4B&	0.021	&-0.098(0.64)	\\
ZNF737&	-0.038	&&&0.007(0.19)	\\
\hline
\end{tabular}
\end{table}

\end{document}